\documentclass[10pt, conference]{IEEEtran}
\IEEEoverridecommandlockouts
\usepackage{cite}
\usepackage{algorithmic}
\usepackage{graphicx}
\usepackage{textcomp}
\usepackage{xcolor}
\usepackage{mathrsfs}
\usepackage{amsfonts,amssymb} 
\usepackage{amsmath,amssymb,amsthm,lipsum}
\usepackage{algorithm}
\usepackage{algorithmic}
\floatname{algorithm}{Network}
\usepackage{bm}
\usepackage{CJK}
\usepackage{enumerate}
\usepackage{diagbox}
\usepackage{float}
\usepackage{caption}
\usepackage{cite}
\usepackage{bbm}
\usepackage{subfigure}
\usepackage{enumitem}
\usepackage{setspace}
\usepackage{bbding}
\usepackage{pifont}
\usepackage{wasysym}
\usepackage{threeparttable}
\usepackage{titlesec}
\usepackage{geometry}

\def\BibTeX{{\rm B\kern-.05em{\sc i\kern-.025em b}\kern-.08em
    T\kern-.1667em\lower.7ex\hbox{E}\kern-.125emX}}
\begin{document}

\title{Sparsity-Based Channel Estimation Exploiting Deep Unrolling for Downlink Massive MIMO\\

\thanks{This work was supported by the National Natural Science Foundation of China (62371053), U.S. National Science Foundation (2136202).}
}

\author{
\IEEEauthorblockN{
An Chen\IEEEauthorrefmark{1},
Wenbo Xu\IEEEauthorrefmark{1},
Liyang Lu\IEEEauthorrefmark{2}, and
Yue Wang\IEEEauthorrefmark{3}}
\IEEEauthorblockA{\IEEEauthorrefmark{1}Key laboratory of Universal Wireless Communications, Beijing University of Posts and Telecommunications, \\
Beijing, P. R. China, 100876}
\IEEEauthorblockA{\IEEEauthorrefmark{2}Department of Electronic Engineering, Tsinghua University, Beijing 100084, China}
\IEEEauthorblockA{\IEEEauthorrefmark{3}Department of Computer Science, Georgia State University, Atlanta, GA 30303, USA}
\IEEEauthorblockA{Email: AnChen@bupt.edu.cn}}

\maketitle

\begin{abstract}
Massive multiple-input multiple-output (MIMO) enjoys great advantage in 5G wireless communication systems owing to its spectrum and energy efficiency. However, hundreds of antennas require large volumes of pilot overhead to guarantee reliable channel estimation in FDD massive MIMO system. Compressive sensing (CS) has been applied for channel estimation by exploiting the inherent sparse structure of massive MIMO channel but suffer from high complexity. To overcome this challenge, this paper develops a hybrid channel estimation scheme by integrating the model-driven CS and data-driven deep unrolling technique. The proposed scheme consists of a coarse estimation part and a fine correction part to respectively exploit the inter- and intra-frame sparsities of channels to greatly reduce the pilot overhead. Theoretical result is provided to indicate the convergence of the fine correction and coarse estimation net. Simulation results are provided to verify that our scheme can estimate MIMO channels with low pilot overhead while guaranteeing estimation accuracy with relatively low complexity.
\end{abstract}

\begin{IEEEkeywords}
Channel estimation, compressive sensing, deep unrolling, massive MIMO.
\end{IEEEkeywords}

\section{Introduction}
Massive MIMO is the key technology in the next-generation wireless communication system to accommodate a broad range of facilities of varying specifications \cite{58}. To enable stable wireless transmission, the channel state information (CSI) in terms of the channel matrix needs to be acquired via channel estimation. Traditional pilot-based channel estimation methods \cite{5,41} rely on sophisticated pilot sequences to guarantee accurate channel estimation performance, which require no prior information about the channel and the estimation is simple to implement. However, the large number of antenna results in enlarged pilot sequences, which involves a large amount of channel coefficients to estimate and causes unaffordable pilot overhead cost \cite{2}. 

In order to efficiently obtain accurate CSI estimation while reducing the pilot overhead, CS has been advocated for CSI estimation in mmWave MIMO systems by leveraging sparsity of MIMO channel in angular domain \cite{9,10,45}. Such sparsity stems from the fact that sparse multipaths propagate between base station (BS) and user equipments (UEs) with limited angular directionalities of massive MIMO channels \cite{45,53,57}. Furthermore, there exist two different types of angular sparsities in massive MIMO channels. The first one is the intra-frame sparsity of MIMO channel within one frame which introduces block sparse structure in MIMO channel matrix \cite{10,56}. Such sparsity is caused by the multipaths that exhibit a limited number of angles of departure (AoDs) at BS, but rich number of angles of arrival (AoAs) at UEs \cite{50}. The second type is inter-frame sparsity, which is further viewed in small-scale and large-scale, because some common AoDs are shared by mobile UEs in different numbers of consecutive frames. The small-scale inter-frame sparsity \cite{9} is measured between two adjacent frames with large number of shared AoDs, while the large-scale inter-frame sparsity \cite{10,55,56} among multiple successive frames with small number of shared AoDs. Although the exploitation of these sparsities can bring advantages in channel estimation performance, the aforementioned CS-based methods are often suffocated by the following two drawbacks: high time complexity to converge and separate considerations on the small-scale and large-scale sparsities. Therefore, it is highly desirable to develop advanced channel estimators with better performance-complexity trade-offs.

Recently, deep unrolling has been shown as an effective solution for many wireless communication problems inspired by its theoretical interpretability and low complexity during prediction \cite{19,61}. For example, the learned-ISTA-coupling weight and support selection (LISTA-CPSS) \cite{19} has been proposed with theoretical interpretability and its variants, e.g., LISTA-group sparse (LISTA-GS) \cite{20} and ALISTA-GS \cite{61} are applied for grant free multiuser detection. For deep unrolling based downlink massive MIMO channel estimation, an end-to-end structure is developed by unifying a pilot design network with an unrolled channel estimation network to improve channel estimation performance \cite{16}. The angular domain sparsity in millimeter-wave massive MIMO is exploited in \cite{62,63}, where the traditional model-based methods in compressive sensing are unrolled by deep learning techniques to improve the performance. Those deep unrolling networks are capable of accurately estimating the channel; however, the joint consideration of different types of sparsities of massive MIMO channel are often ignored by these networks and thus leaves great space for improving the channel estimation performance.

The contributions of this paper are summarized as follows:  First, a two-stage channel estimation scheme is proposed, by exploiting the large-scale inter-frame sparsity in the coarse stage and the intra- and small-scale inter-frame sparsity in the fine stage, respectively. Second, in order to reduce time complexity while maintaining the channel estimation performance, deep unrolling technique is employed with the proposed two-stage scheme. Third, simulation results are provided to verify that our scheme can estimate channel with less pilot overhead and achieve better trade-off between channel estimation performance and time complexity compared with baseline schemes.

The rest of this paper is organized as follows: In Section ${\rm\uppercase\expandafter{\romannumeral2}}$, the system model and the problem formulation are described. In Section ${\rm\uppercase\expandafter{\romannumeral3}}$, we present the two-stage channel estimation scheme to reconstruct the MIMO channels. In Section ${\rm\uppercase\expandafter{\romannumeral4}}$, simulation results are presented to indicate the superiority of the proposed scheme over the baseline schemes, followed by conclusions in Section ${\rm\uppercase\expandafter{\romannumeral5}}$.

\section{SYSTEM MODEL AND PROBLEM DESCRIPTIONS}

\subsection{Channel Model}

In this paper, a point-to-point single-user downlink massive MIMO system is considered, with $M$ transmit antennas at BS side and $N$ receive antennas at UE side ($N \ll M$).  We adopt a block-fading channel model in which pilot signal of length $T$ is utilized by BS to perform channel estimation. The observed channel in the $i$-th frame can be expressed as follows:
\begin{equation}\label{CSmodel}
\mathbf{Y}^{[i]}= \mathbf{H}^{[i]}\mathbf{X} + \mathbf{N}^{[i]}, 
\end{equation}
where $\mathbf{H}^{[i]}\in \mathbb{C}^{N\times M}$ represents the downlink channel from BS to UE, $\mathbf{X}\in \mathbb{C}^{M\times T}$ denotes the transmitted pilot matrix, $Tr(\mathbf{X}^H\mathbf{X})=T$, 
and $\mathbf{N}^{[i]}$ is the noise matrix whose elements are i.i.d. complex Gaussian random variables with zero mean and variance $\delta^2$.

In practical scenario, the number of multipaths in MIMO channel is limited due to the limited scatterings \cite{9,57}, resulting in a sparse structure after applying a virtual angular domain transformation on the MIMO channel in (\ref{CSmodel}). The virtual angular domain transformation with fixed virtual receive and transmit directions is expressed as follows \cite{53,57}:
\begin{equation}\label{ordinary model}
\normalsize
\mathbf{H}^{[i]}=\mathbf{U}\mathbf{\tilde{H}}^{[i]}\mathbf{V}^H,
\end{equation}
where $ \mathbf{\tilde{H}}^{[i]} \in \mathbb{C}^{N\times M} $ is the virtual angular domain representation of $\mathbf{H}^{[i]}$, $\mathbf{U}\in \mathbb{C}^{N\times N}$ and $\mathbf{V}\in \mathbb{C}^{M\times M}$ denote unitary spatial Fourier transform matrices for the angular domain transformation at the UE side and the BS side, respectively. The nonzero entries in $ \mathbf{\tilde{H}}^{[i]} $ denote the complex gains of the corresponding paths. Due to the limited scatterings in MIMO channel \cite{57}, there only exists a limited number of nonzero entries in (\ref{ordinary model}). Hence, CS-based methods can be applied to exploit the sparsity of the angular domain channel matrix.

\subsection{Sparsity Model}
\begin{figure}
  \centering
  \includegraphics[scale=0.4]{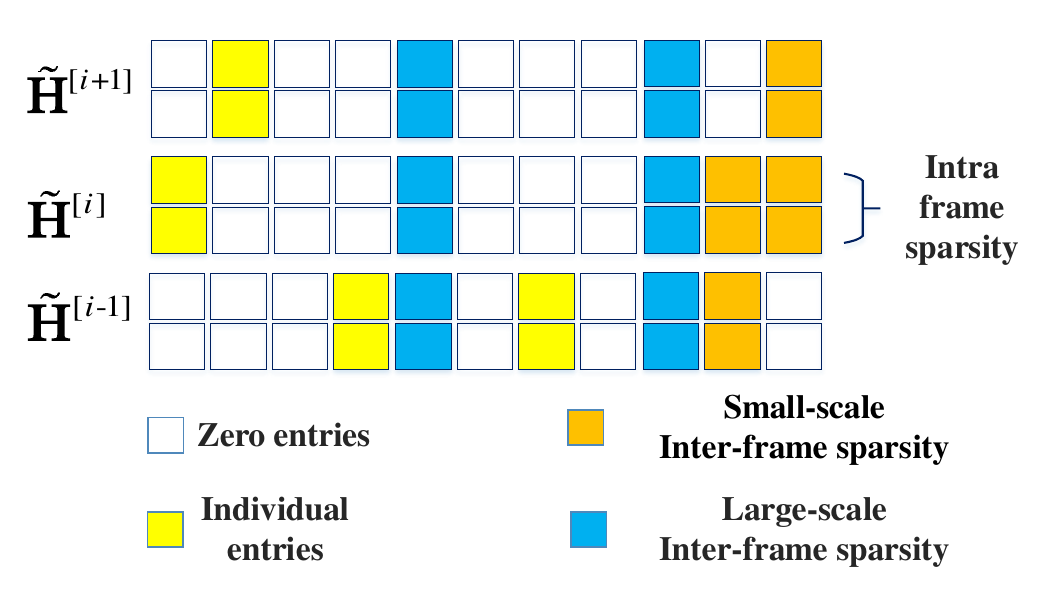}\\
  \caption{Illustration of angular domain channel model. Different colours in this figure represent different kinds of spatial paths affected by different scatterers.}\label{sparse_signal}
\vspace{-2.0em}
\end{figure}

According to \cite{9,10,56}, BS is located at high elevation with limited local scatterings, but UE is usually located at low elevation with rich local scatterings, resulting in limited spatial paths which depart from BS but rich spatial paths which arrive at UE. When UE moves from one place to another place, some common scatterers are shared among its successive receiving frames, which brings the correlation in angular domain.
From the practical observations of previous work \cite{9,10,56}, the channel matrices in angular domain have the following characteristics:

\begin{itemize}[leftmargin=1pt, labelsep=4pt,itemindent=20pt, listparindent=10pt]
\item \textbf{ Intra-frame sparsity:} In the downlink massive MIMO system, all the spatial paths depart from BS in limited directions and arrive at UE in omni-directions \cite{9}. Accordingly, each row of the channel matrix $ \mathbf{\tilde{H}}^{[i]} $ bears the same support set \cite{10,56}, i.e., $ supp(\mathbf{\tilde{H}}^{[i]}_{(1,:)})=\{j \ | \ \mathbf{\tilde{H}}^{[i]}_{(1,j)} \neq 0\}=\cdots=supp(\mathbf{\tilde{H}}^{[i]}_{(N,:)}) \triangleq \mathbf{\Gamma}^{[i]}$. The example of $N=2$ is depicted for illustration in Fig. \ref{sparse_signal}. The cardinality is generally upper bounded $|\mathbf{\Gamma}^{[i]}| \leq \overline{s}$. Furthermore, the nonzero elements in $\mathbf{\tilde{H}}^{[i]}$ are assumed to be i.i.d. complex Gaussian distributed with zero mean and unit variance \cite{9}. 

\item \textbf{ Inter-frame sparsity:} 
Practically, massive MIMO channels for consecutive frames are usually intercorrelated because of the shared scatterers at BS, which result in shared AoDs in consecutive frames \cite{9,10,56}.
The first type is the small-scale inter-frame sparsity that models the correlation of AoDs between two adjacent frames \cite{9}, which can be represented by the shared support indices:
\begin{equation}\label{small inter common support set}
\overline{s} > | \mathbf{\Gamma}^{[i-1]} \cap \mathbf{\Gamma}^{[i]}| \geq s_c,
\end{equation}
where $s_c$ and $\overline{s}$ denote the upper bound and lower bound of such intersection set. The orange boxes depicted in Fig. \ref{sparse_signal} illustrate this kind of correlation.

The second type is the large-scale inter-frame sparsity which models the correlation of AoDs among $L$ consecutive frames \cite{10,56}, which can be represented by the common row support set:
\begin{equation}\label{inter common support set}
\mathbf{\Gamma}^s = \cap_{i=1}^{L} \mathbf{\Gamma}^{[i]},
\end{equation}
The cardinality $|\mathbf{\Gamma}^s|=S$. For a clear demonstration of (\ref{inter common support set}), the blue boxes are depicted in Fig. \ref{sparse_signal} to illustrate this kind of correlation with $L=3$ for illustration.
\end{itemize}

\subsection{Problem Formulation}
Based on (\ref{ordinary model}), the model (\ref{CSmodel}) is rewritten as follows:
\begin{equation}\label{angular burst model}
\underbrace{(\mathbf{Y}^{[i]})^H \mathbf{U}}_{\mathbf{Z}^{[i]}}= \underbrace{ (\mathbf{X}^H \mathbf{V})}_{\mathbf{\Phi}} \underbrace{(\mathbf{\tilde{H}}^{[i]})^H}_{\mathbf{S}^{[i]}} + \underbrace{(\mathbf{N}^{[i]})^H \mathbf{U}}_{\mathbf{W}^{[i]}},
\end{equation}
where $\mathbf{Z}^{[i]}\in \mathbb{C}^{T\times N}$, $\mathbf{\Phi}\in \mathbb{C}^{T\times M}$, $\mathbf{W}^{[i]}\in \mathbb{C}^{T\times N}$ and $\mathbf{S}^{[i]}\in \mathbb{C}^{M\times N}$. Note that $\mathbf{S}^{[i]}$  is a block sparse matrix with only $|\mathbf{\Gamma}^{[i]}|$ nonzero rows, representing the intra-frame sparsity at BS side.

When a UE with $N$ antennas receives $L$ consecutive frames from BS, we concatenate the received signals as $\mathbf{R}=[\mathbf{Z}^{[1]},\cdots,\mathbf{Z}^{[L]}]$ in the form of:
\begin{equation}\label{overall angular burst model}
\mathbf{R}=\mathbf{\Phi} \mathbf{G} + \mathbf{N},
\end{equation}
where $\mathbf{G}=[\mathbf{S}^{[1]},\cdots,\mathbf{S}^{[L]}]$ and $\mathbf{N}=[\mathbf{W}^{[1]},\cdots,\mathbf{W}^{[L]}]$ denote the concatenated row-sparsity channel matrices and the noise across $L$ consecutive frames, respectively.

The optimization objective of massive MIMO channel estimation is modelled as follows:
\begin{equation}\label{overall objective}
\min_{\mathbf{G} \in \mathbb{C}^{M \times NL}}\frac{1}{2}||\mathbf{R}-\mathbf{\Phi}\mathbf{G}||_F^2 + \xi(\mathbf{G}),
\end{equation}
where $\xi(\mathbf{G})$ means the sparse regularizer on the concatenated row-sparse channel matrix $\mathbf{G}$.

However, it is hard to directly reconstruct $\mathbf{G}$ via the above optimization problem since no reasonable sparse regularizer $\xi(\cdot)$ can effectually capture the complicated sparse structure induced by the intra-frame sparsity and the two kinds of inter-frame sparsity simultaneously.  To solve the objective (\ref{overall objective}), a two-stage structure is proposed in this work which consists of the coarse estimation part and the fine correction part, where the large-scale and the small-scale inter-frame sparsities are exploited by these two parts, respectively.

By considering the sparse structure induced by the large-scale inter-frame sparsity among consecutive frames in (\ref{overall angular burst model}), we propose the coarse estimation part to estimate the corresponding coefficients, and its objective can be written as:
\begin{equation}\label{coarse objective}
\begin{aligned}
&\min_{\mathbf{G} \in \mathbb{C}^{M \times NL}}\frac{1}{2}||\mathbf{R}-\mathbf{\Phi}\mathbf{G}||_F^2 + \alpha \sum_{j=1}^M ||\mathbf{G}_{(j,:)}||_{2},
\end{aligned}
\end{equation}
which is a standard $l_{2,1}$ minimization problem and $\alpha$ stands for the regularization parameter.

However, the above objective cannot accurately reconstruct MIMO channel since the status of the entries in some partial-nonzero columns (such as the columns with yellow and orange entries in Fig. \ref{sparse_signal}) cannot be determined through the simple $l_{2,1}$ minimization in (\ref{coarse objective}). Therefore, in fine correction part, a weighted $\ell_{2,1}$ minimization objective is designed based on the decoupled channel model (\ref{angular burst model}), which can be written as follows:
\begin{equation}\label{weighted fine objective}
\min_{\mathbf{S}^{[i]}\in \mathbb{C}^{M\times N}}\frac{1}{2}||\mathbf{Z}^{[i]}-\mathbf{\Phi}\mathbf{S}^{[i]}||_F^2 + \lambda \sum_{j=1}^M \omega_j ||\mathbf{S}_{(j,:)}^{[i]}||_{2}
\end{equation}
with
\begin{equation}\nonumber
\begin{aligned}
&\ \omega_{j}=
\begin{cases}
1, & j \notin \hat{\mathbf{\Gamma}}^{[i-1]}=supp(\mathbf{\hat{S}}^{[i-1]})=\lbrace j \ | \ ||\mathbf{\hat{S}}^{[i-1]}_{(j,:)}||_2 \neq 0\rbrace,\\
\omega, & j \in \hat{\mathbf{\Gamma}}^{[i-1]},
\end{cases}
\end{aligned}
\end{equation}
In this objective, $\omega$ is a training parameter learned from massive data, which reflects the degree of small-scale correlation between the channels in the $i$-th frame and the $i-1$-th frame. Meanwhile,  the intra-frame sparsity is considered by regularizing channel $\mathbf{S}^{[i]}$ through weighted $l_{2,1}$ norm. 

\section{The description of proposed scheme}
\addtolength{\topmargin}{0.08in}
In this section, we describe the proposed two-stage channel estimation scheme to solve the objective (\ref{coarse objective}) and (\ref{weighted fine objective}), by exploiting large-scale and small-scale inter-frame sparsities, respectively.

\begin{figure*}
	\centering
   \includegraphics[scale=0.38]{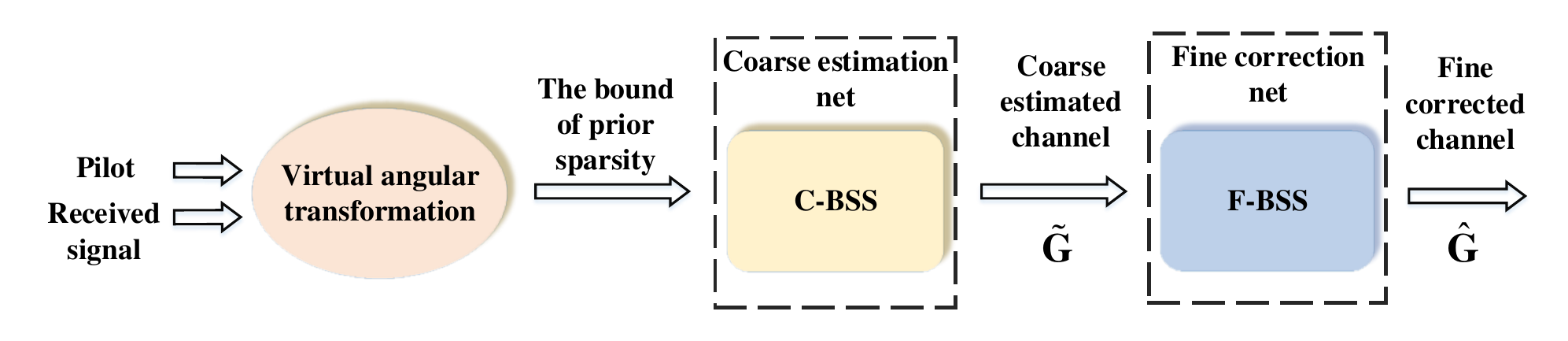}
	\caption{The whole network structure of the proposed scheme.}
	\label{whole network}
	\vspace{-5pt}
\end{figure*}
\subsection{Coarse Estimation Net}
To facilitate the network design, we first give the real-valued counterpart of (\ref{overall angular burst model}):
\begin{equation}\label{coarse angular real model}
\begin{aligned}
\bar{\mathbf{R}}&=\bar{\mathbf{\Phi}} \bar{\mathbf{G}}+ \bar{\mathbf{N}}\\
&=\begin{bmatrix} \mathbbm{R}(\mathbf{\Phi}) & -\mathbbm{I}(\mathbf{\Phi})  \\ \mathbbm{I}(\mathbf{\Phi}) & \mathbbm{R}(\mathbf{\Phi}) \end{bmatrix}    \begin{bmatrix} \mathbbm{R}(\mathbf{G})  \\ \mathbbm{I}(\mathbf{G}) \end{bmatrix} +  \begin{bmatrix} \mathbbm{R}(\mathbf{N})  \\ \mathbbm{I}(\mathbf{N}) \end{bmatrix},
\end{aligned}
\end{equation}
where $\bar{\mathbf{R}}$, $\bar{\mathbf{\Phi}}$, $\bar{\mathbf{G}}$ and $\bar{\mathbf{N}}$ denote the real-valued version of received signal, measurement matrix, actual channel matrix and noise, respectively.

Inspired by the traditional ISTA-based algorithms in \cite{21} to solve the $l_{2,1}$ minimization problem, we develop the coarse estimation net in Network \ref{alg:coarse estimation net} to solve the objective (\ref{coarse objective}) of coarse estimation part.
The coarse estimation net can be viewed as an extended version of deep neural network (DNN), where parameters $\mathbf{W}^l_{e}$ and $\theta_e^l$ stand for training weights in the $l$-th layer and $\gamma_{\theta_e^l}(\cdot, \mathbf{\Omega}^l_e)$ represents activation function with additional input $\mathbf{\Omega}^l_e$. Specifically, this activation function is named as block thresholding function with support selection (BSS) by generalizing SS function in \cite{19}. It is defined as follows:
\begin{equation}\label{soft thre support selection function}
\gamma_{\mathbf{\theta}^l_e}(\mathbf{V}_{(j,:)}, \mathbf{\Omega}^{l}_e) = \frac{\mathbf{V}_{(j,:)}}{||\mathbf{V}_{(j,:)}||_2} \eta_{\mathbf{\theta}^l_e }(\mathbf{V}_{(j,:)}, \mathbf{\Omega}^{l}_e),
\end{equation}
where
\begin{equation}\label{right part of soft thre support selection function}
\begin{aligned}
&\eta_{\mathbf{\theta}^l_e}(\mathbf{V}_{(j,:)}, \mathbf{\Omega}^{l}_e) = \\
&\left\{
\begin{aligned}
&||\mathbf{V}_{(j,:)}||_2, & {\rm if}\ & ||\mathbf{V}_{(j,:)}||_2 > \mathbf{\theta}^l_e, j \in \mathbf{\Omega}^{l}_e, \\
&||\mathbf{V}_{(j,:)}||_2 - \mathbf{\theta}^l_e ,& {\rm if}\ &  ||\mathbf{V}_{(j,:)}||_2 > \mathbf{\theta}^l_e, j \notin \mathbf{\Omega}^{l}_e, \\
&0,& {\rm if}\ & ||\mathbf{V}_{(j,:)}||_2 \leq \mathbf{\theta}^l_e.\\
\end{aligned}
\right.\\
\end{aligned}
\end{equation}
\begin{algorithm}[t]
	\renewcommand{\algorithmicrequire}{\textbf{Input:}}
	\renewcommand{\algorithmicensure}{\textbf{Output:}}
	\caption{Coarse Estimation Net}
	\label{alg:coarse estimation net}
	\begin{algorithmic}[1]
		\REQUIRE $\bar{\mathbf{R}}$, $\bar{\mathbf{\Phi}}$, $S$, \rm the number of layers$\ L_{e}$
		\ENSURE $\tilde{\mathbf{G}}$
        \STATE $\mathbf{Initialization:}$ $l=1$, $\mathbf{G}^0=\mathbf{0}_{M\times LN}$
        \WHILE {$l\leq L_{e}$ }
		\STATE $\mathbf{G}^l=\gamma_{\theta_e^l}(\mathbf{G}^{l-1} + \mathbf{W}^l_{e}(\bar{\mathbf{R}}-\bar{\mathbf{\Phi}}\mathbf{G}^{l-1}),\mathbf{\Omega}^l_e)$
		\STATE $l=l+1$
        \ENDWHILE
		\STATE \textbf{return}  $\tilde{\mathbf{G}}=\mathbf{G}^{L_{e}}=[\tilde{\mathbf{S}}^{[1]},\cdots,\tilde{\mathbf{S}}^{[L]}]$
	\end{algorithmic}
\vspace{-0.1cm}
\end{algorithm}

Such thresholding function shows advantages on recoverability and convergence by exploiting the additional information, i.e., the selected support set $\mathbf{\Omega}^l_e$, which relies on the prior sparsity information and inexact reconstruction of the current estimation. The set $\mathbf{\Omega}^{l}_e$ includes the indices of the elements with the largest $p_{_l}\%$ magnitudes in current estimations $\lbrace ||\mathbf{V}_{(j,:)}||_2\ |\  j=1,\cdots,M\rbrace$:
\begin{equation}\label{support selection set}
\mathbf{\Omega}^{l}_e = \lbrace j_i,\cdots,j_{\lfloor p_lM \rfloor}\ | \ ||\mathbf{V}_{(j_1,:)}||_2 \geq \cdots \geq ||\mathbf{V}_{(j_M,:)}||_2 \rbrace.
\end{equation} 
The following equation is proposed to calculate the value of $p_{_l}$:
\begin{equation}\label{ss thresholding}
p_l = (\frac{p_{min}}{M} + \frac{p_{max}-p_{min}}{M(L-1)} (l-1))\times100\%, l=1,\cdots,L_e.
\end{equation}
The above equation indicates that $p_l \in [p_{min}/M, p_{max}/M]$, where $p_{max}$ and $p_{min}$ denote the upper bound and the lower bound of the cardinality of $\mathbf{\Gamma}^s$, respectively. 

In a word, the coarse estimation net with BSS (C-BSS) is proposed, which relies on the prior sparsity information to solve the objective (\ref{coarse objective}). For more details about the choice of sparsity bound information $(p_{min},p_{max})$, please refer to the arxiv version of our paper$\footnote{The arxiv version of our paper has been uploaded to http://arxiv.org/abs/2210.17212\label{adress}}$.

\subsection{Fine correction Net}
\addtolength{\topmargin}{-0.04in}
\begin{algorithm} [t]
	\renewcommand{\algorithmicrequire}{\textbf{Input:}}
	\renewcommand{\algorithmicensure}{\textbf{Output:}}
	\caption{Fine Correction Net}
	\label{alg:fine correction net}
	\begin{algorithmic}[1]
		\REQUIRE  \rm The number of layers$\ L_{c}$, \rm the number of frames $L$, $\bar{\mathbf{\Phi}}$, $\lbrace \bar{\mathbf{Z}}^{[i]}| i=1,\cdots,L \rbrace$ and coarsely estimated channels $\lbrace \tilde{\mathbf{S}}^{[i]}| i=1,\cdots,L \rbrace$.
		\ENSURE $\hat{\mathbf{G}}$
		\WHILE {$i\leq L$ }
        \STATE $\mathbf{Initialization:}$ $l=1$, $\mathbf{S}^0=\tilde{\mathbf{S}}^{[i]}$, $\hat{\mathbf{\Gamma}}^{[0]}=\emptyset$ and $\hat{\mathbf{\Gamma}}^{[i-1]}={supp(\hat{\mathbf{S}}^{[i-1]})}$\\
        \WHILE {$l\leq  L_{c}$}
		\STATE $\mathbf{S}^l= \gamma_{\mathbf{\theta}_c^l \bm{\omega}} (\mathbf{S}^{l-1} + \mathbf{W}^l_{c}(\bar{\mathbf{Z}}^{[i]}-\bar{\mathbf{\Phi}}\mathbf{S}^{l-1}),\mathbf{\Omega}^l_c)$
		\STATE $\bm{\omega}=[\omega_1,\cdots,\omega_M]$, where $\omega_j = 
				\left\{
				\begin{aligned}
				&\omega : & \ j \in \hat{\mathbf{\Gamma}}^{[i-1]} \\
				& 1 : & \  j \notin \hat{\mathbf{\Gamma}}^{[i-1]}\\
				\end{aligned}
				\right.$
		\STATE $l=l+1$
        \ENDWHILE
		\STATE $\hat{\mathbf{S}}^{[i]} = \mathbf{S}^{ L_{c}}$
		\ENDWHILE
		\STATE \textbf{return}  $\hat{\mathbf{G}}=[\hat{\mathbf{S}}^{[1]},\cdots,\hat{\mathbf{S}}^{[L]}] $
	\end{algorithmic}
\end{algorithm}
We first give the real-valued counterpart of (\ref{weighted fine objective}): 
\begin{equation}\label{fine angular real model}
\begin{aligned}
\bar{\mathbf{Z}}^{[i]}&=\bar{\mathbf{\Phi}} \bar{\mathbf{S}}^{[i]}+\bar{\mathbf{W}}^{[i]}\\
&=\begin{bmatrix} \mathbbm{R}(\mathbf{\Phi}) & -\mathbbm{I}(\mathbf{\Phi})  \\ \mathbbm{I}(\mathbf{\Phi}) & \mathbbm{R}(\mathbf{\Phi}) \end{bmatrix}    \begin{bmatrix} \mathbbm{R}(\mathbf{S}^{[i]})  \\ \mathbbm{I}(\mathbf{S}^{[i]}) \end{bmatrix} +  \begin{bmatrix} \mathbbm{R}(\mathbf{W}^{[i]})  \\ \mathbbm{I}(\mathbf{W}^{[i]}) \end{bmatrix},
\end{aligned}
\end{equation}
where $\bar{\mathbf{\Phi}}$, $\bar{\mathbf{Z}}^{[i]}$ and $\bar{\mathbf{S}}^{[i]}$ denote the real-valued version of measurement matrix, the received signal and the channel in the $i$-th frame, respectively. 

The unrolled structure of fine correction net is proposed in Network \ref{alg:fine correction net} to solve the objective (\ref{weighted fine objective}) of fine correction part. The parameters $\mathbf{W}^l_{c}$ and $\theta_c^l$ stand for training weights in the $l$-th layer. Moreover, a new activation function (namely generalized BSS) based on (\ref{soft thre support selection function}) is proposed to accommodate small-scale inter-frame sparsity, which is defined as follows:
\begin{equation}\label{soft weighted thre support selection function}
\gamma_{\mathbf{\theta}_c^l  \omega_j}(\mathbf{V}_{(j,:)},\mathbf{\Omega}^{l}_c) = \frac{\mathbf{V}_{(j,:)}}{||\mathbf{V}_{(j,:)}||_2} \eta_{\mathbf{\theta}_c^l  \omega_j}(\mathbf{V}_{(j,:)},\mathbf{\Omega}^{l}_c)
\end{equation}
\begin{equation}\nonumber
\begin{aligned}
&\eta_{\mathbf{\theta}_c^l \omega_j}(\mathbf{V}_{(j,:)},\mathbf{\Omega}^{l}_c) = \\
&\left\{
\begin{aligned}
&||\mathbf{V}_{(j,:)}||_2, & {\rm if}\ & ||\mathbf{V}_{(j,:)}||_2 > \mathbf{\theta}_c^l, j \in \mathbf{\Omega}^{l}_c, \\
&||\mathbf{V}_{(j,:)}||_2 - \mathbf{\theta}_c^l \omega_j,& {\rm if}\ &  ||\mathbf{V}_{(j,:)}||_2 > \mathbf{\theta}_c^l \omega_j, j \notin \mathbf{\Omega}^{l}_c, \\
&0,& {\rm if}\ & ||\mathbf{V}_{(j,:)}||_2 \leq \mathbf{\theta}_c^l\omega_j.\\
\end{aligned}
\right.\\
\end{aligned}
\end{equation}

\begin{table*}
\renewcommand{\arraystretch}{1.1}
\caption{Information utilized in the baseline schemes.}
\label{existing schemes setting}
\centering
\begin{threeparttable}
\begin{tabular}{|m{32mm}|m{32mm}|m{32mm}|m{32mm}|m{32mm}|}
\hline
\textbf{Schemes} & \textbf{Prior Sparsity Bound} $(\bar{s},s_c)$ &\textbf{Intra-Frame Sparsity} &\textbf{Small-Scale Inter-Frame Sparsity} &\textbf{Large-Scale Inter-Frame Sparsity}\\
\hline
M-SP \cite{9} & $\checked$ & $\checked$ & $\checked$  & $\times$ \\
\hline
Enhanced MMV \tnote{2}  \cite{10} & $\checked$ & $\checked$ & $\times$  & $\checked$ \\
\hline
LISTA-CPSS \cite{19} & $\checked$ & $\times$ & $\times$  & $\times$ \\
\hline
LISTA-GS \cite{20} & $\times$ & $\checked$ & $\times$  & $\times$ \\
\hline
\end{tabular}

\begin{tablenotes}
        \footnotesize
        \item[2] BCD-MMV algorithm in \cite{21} is utilized to solve the minimization problem in this scheme for fair comparison since our proposed fine correction net also originates from BCD-MMV. 
\end{tablenotes}
\vspace{-6mm}
\end{threeparttable}

\end{table*}
The difference between (\ref{soft weighted thre support selection function}) and (\ref{soft thre support selection function}) lies in the exploitation of small-scale inter-frame sparsity, where the trainable weight $\omega$ is optimized during the training process to indicate the degree of small-scale correlation between two adjacent frames.

Moreover, the calculation of $\mathbf{\Omega}^l_c$ is also based on (\ref{support selection set}), where $p_{max}=\overline{s}$ and $p_{min}=s_c$ denote the upper bound and the lower bound of the cardinality of $\mathbf{\Gamma}^{[i]}$, respectively. 

Based on the above discussions, our proposed two-stage structure is depicted in Fig. \ref{whole network}, which consists of coarse estimation net and fine correction net with BSS (C-F-BSS). The detailed designs of C-F-BSS and the complexity analysis can be found in the journal version of our paper$\textsuperscript{1}$.

\subsection{Convergence analysis}
Inspired by \cite{52}, we provide the convergence analysis of fine correction net F-BSS. This theoretical result can also be applied to C-BSS since the structure of F-BSS is the generalized version of C-BSS by considering small-scale inter-frame sparsity.

\emph {Theorem 1 (Convergence of F-BSS)}: Given $\lbrace \theta_{c}^l, \mathbf{W}^l_{c} \rbrace_{l=0}^{\infty}$ and $\mathbf{S}^0 = \mathbf{0}$, let $\lbrace \mathbf{S}^l \rbrace_{l=1}^{\infty}$ denote the $l$-th layer of F-BSS. With the same definition (\emph{Definition 1} and \emph{Definition 2} in \cite{52}) and assuming $\{({{\bar{\mathbf{S}}}^{[i]}},{{\bar{\mathbf{W}}}^{[i]}})\ |\ ||{{\bar{\mathbf{S}}}^{[i]}}[j,:]|{{|}_{2}}\le \beta ,\forall j,||{{\bar{\mathbf{S}}}^{[i]}}|{{|}_{2,0}}\le \bar{s},||{{\bar{\mathbf{W}}}^{[i]}}||_{F}\le \sigma \}$. The upper bound of the estimation error for $ \mathbf{S}^l$ and ground truth ${{\bar{\mathbf{S}}}^{[i]}}$ is given by:
\begin{equation}
\begin{aligned}
&||{{\mathbf{S}}^{l}}-{{\bar{\mathbf{S}}}^{[i]}}|{{|}_{F}}\le \bar{s}\rho\prod\limits_{k=0}^{l-1} e ^{-c_{ss}^k} + C_{ss}\sigma,
\end{aligned}
\end{equation}
with 
\begin{equation}\label{fine addon}
\begin{aligned}
&c_{ss}^k=-\log (\tilde{\mu}((c(\omega+1) +2)\bar{s} - 2s_c - 1 - \inf \limits_{({{\bar{\mathbf{S}}}^{[i]}},{{\bar{\mathbf{W}}}^{[i]}})} |\pi^{k}_c|)),\\
&C_{ss} = ((c(\omega+1) +2)\bar{s} - 2s_c) C_W \sum\limits_{k=\text{1}}^{l-1} \prod\limits_{t=k}^{l-1} e^{-c_{ss}^k}, \\
&\pi_{c}^{k} = \lbrace i \ | \ {{\mathbf{S}}^{k+1}}_{(i,:)} \neq \mathbf{0}, i \in \mathbf{\Omega }_{c}^{k}\rbrace, \ c \in 0 \cup [1,M/\bar{s}],
\end{aligned}
\end{equation}
where $c$ in (\ref{fine addon}) is a scaling parameter that indicates the cardinality of $\hat{\mathbf{\Gamma}}^{[i-1]}$ in its worst case, i.e., all the rows of $\mathbf{\hat{S}}^{[i-1]}$ except the ones that can be proved as zero rows are reconstructed as nonzero rows, thus we have $|\hat{\mathbf{\Gamma}}^{[i-1]}|=c\bar{s}>\bar{s}$.  The definitions of $C_W$, $\tilde{\mu}$ and the complete proof of Theorem 1 can be found in the journal version of our paper. 

 This bound ensures the convergence of F-BSS network as the number of layers goes into infinity. When $c=0$ and $s_c = 0$, it reduces to the upper bound of C-BSS.  Moreover, when the small-scale inter-frame sparsity is strong and the prior support information $\hat{\mathbf{\Gamma}}^{[t-1]}$ is accurate, i.e., $s_c = \bar{s}, c=1$ and $\omega \in (0,1)$, it can be concluded that F-BSS can achieve lower error upper bound compared with F-BSS without small-scale inter-frame sparsity.
\section{Simulation results}
In our simulation, the BS and UE are equipped with $M=128$ and $N=2$ antennas, respectively. The angular domain channel coefficients in $\mathbf{G}$ are i.i.d. complex Gaussian with zero mean and unit variance as those in \cite{9}. Suppose that the number of spatial paths from BS to UE is randomly generated as $|\mathbf{\Gamma}^{[i]}| \sim \mathcal{Z}(\overline{s}-3,\overline{s}-1)$ and the number of shared paths between consecutive frames obeys $| \mathbf{\Gamma}^{[i-1]} \cap \mathbf{\Gamma}^{[i]}| \sim \mathcal{Z}(s_c,s_c+1)$, where $\mathcal{Z}(a,b)$ denotes integer uniform distribution between integer $a$ and integer $b$. The number of frames that are jointly considered is $L=7$. The angle of departure is uniformly distributed over $(0, \pi)$, indicating the equal probability of each column in $\mathbf{\tilde{H}}^{[i]}$ to be a nonzero column. The elements in the pilot matrix are i.i.d. and drawn from the uniform distribution $\mathcal{U}(-\sqrt{1/M},\sqrt{1/M})$.

Note that the coarse estimation net and the fine correction net are trained layer by layer as in \cite{19}. At the training stage, the parameters  $\lbrace \mathbf{W}^l_{e}, \theta_e^l \rbrace ^{L_{e}}_{l=1}$ in the coarse estimation net are trained first but are frozen when we train the parameters $\lbrace \mathbf{W}^l_{c}, \theta_c^l, \omega \rbrace ^{L_{c}}_{l=1}$ in the fine correction net. Besides, we find that after the respective training of coarse and fine nets, fine-tuning step with both two nets brings no obvious performance improvements. Thus, we omit fine-tuning step of the proposed scheme for simplicity. Four baseline schemes in Table \ref{existing schemes setting} are provided to demonstrate the superiority of our scheme, where check mark means that such term is included in the corresponding scheme while the cross mark is on the contrary. The number of layers for coarse estimation net $L_{e}$ and fine correction net $L_{c}$ are set as $8$ and $16$, respectively.
The dataset $\lbrace \bar{\mathbf{R}}_i,\bar{\mathbf{G}}_i \rbrace ^{K}_{i=1}$ is generated according to \cite{9,10} and equation (\ref{angular burst model}), (\ref{overall angular burst model}), (\ref{coarse angular real model}) and (\ref{fine angular real model}).  The number of samples in the training, validation and test dataset is $K=20000$, $K=5000$ and $K=1000$, respectively. Additionally, training and validation batch sizes are set as 32 and 100, respectively. We use the adam optimizer and the learing rate is $0.0005$. Both the coarse estimation net and the fine correction net use mean square error (MSE) as loss function. The normalized MSE (NMSE) and the lower bound for achievable spectrum efficiency (ASE) \cite{64} metric are chosen to assess the channel estimation performance which are respectively defined as follows:
\begin{equation}
{\rm NMSE} = \frac{1}{K}\sum_{j=1}^K 10\text{log}_{10}\bigg( \frac{||\bar{\mathbf{G}}_j- \hat{\mathbf{G}}||_F}{||\bar{\mathbf{G}}_{j}||_F}\bigg)
\end{equation}

\begin{equation}
{\rm ASE} = \mathbb{E}\bigg( log_2 \bigg| \mathbf{I}_N + (MN({\rm 10^{NMSE/10}} + \delta^2)^{-1}) \mathbf{H}^{[i]} (\mathbf{H}^{[i]})^H \bigg| \bigg)
\end{equation}

\begin{figure}
\setlength{\belowcaptionskip}{-20pt}
  \centering
	\hspace{-10mm}	
	\begin{minipage}[t]{0.19\textwidth}
	  \includegraphics[scale=0.35]{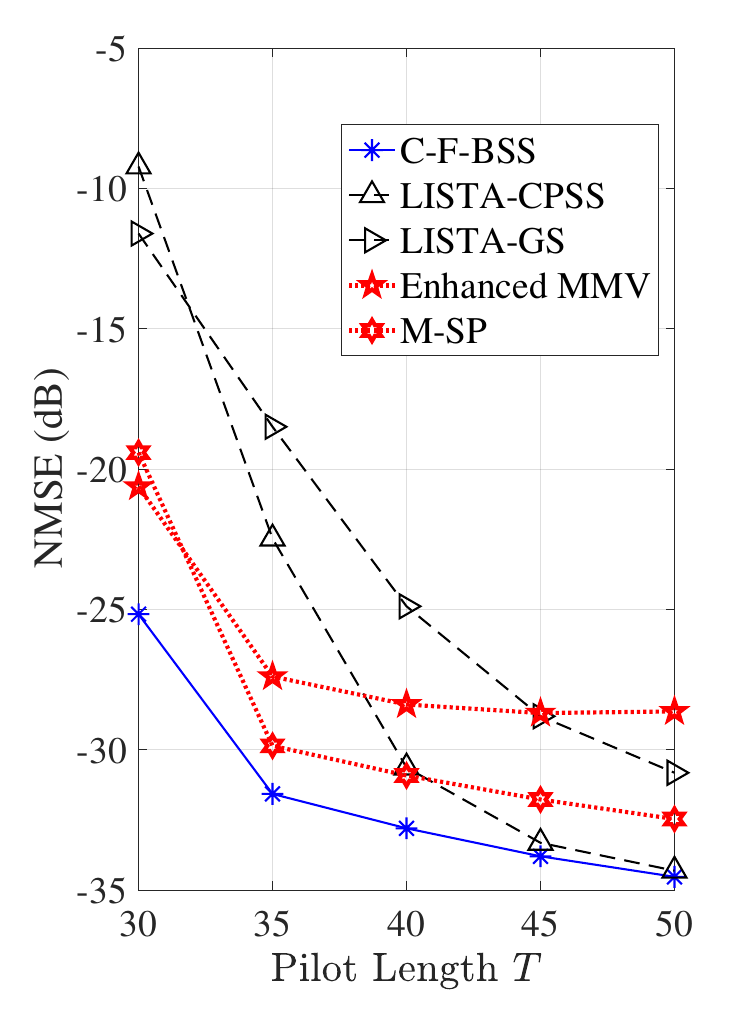}\\
	\end{minipage}
	\hspace{10mm}
	\begin{minipage}[t]{0.19\textwidth}
	  \includegraphics[scale=0.35]{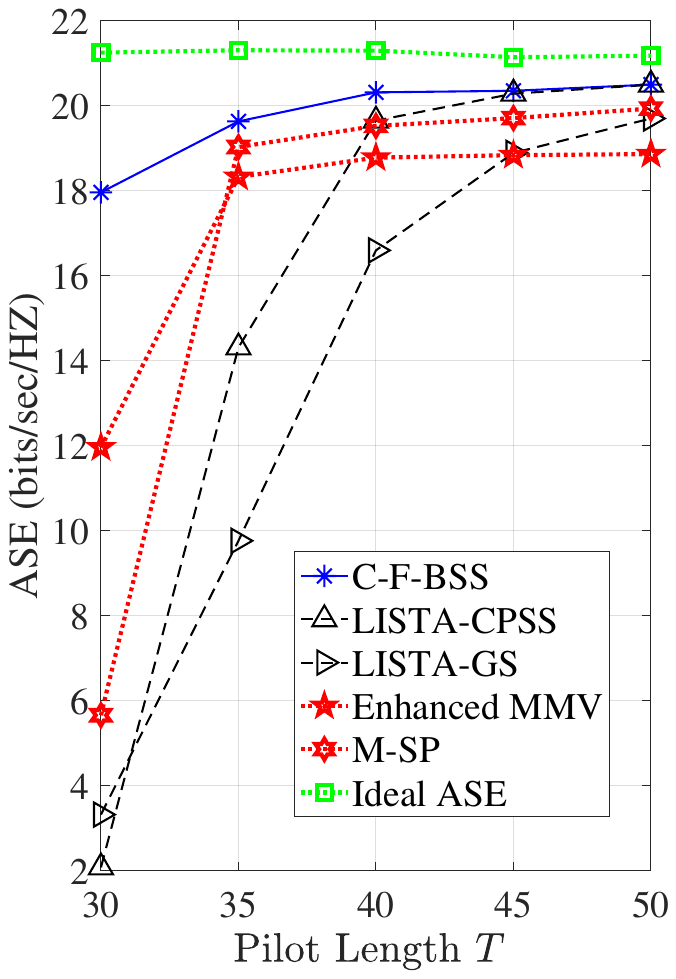}\\
	\end{minipage}
	\vspace{-5mm}
 \caption{NMSE and ASE comparison of estimated channel versus pilot length $T$ under $\overline{s}=15$, $s_c=10$ and $\text{SNR}=30$ dB.}\label{CR}
\end{figure}
Fig. \ref{CR} plots the impact of pilot length $T$ on the NMSE and ASE performance of all schemes, respectively. It is clear that our proposed scheme achieves better channel estimation performance compared with baseline schemes in a wide range of pilot length. This figure also demonstrates that our scheme requires lower pilot length when achieving the same NMSE and ASE performance. Thus, in massive MIMO system, our scheme guarantees robust transmission and saves communication resources by requiring less pilot overhead.
\begin{figure}
\setlength{\belowcaptionskip}{-15pt}
  \centering
	\hspace{-10mm}
	\begin{minipage}[t]{0.19\textwidth}
	  \includegraphics[scale=0.35]{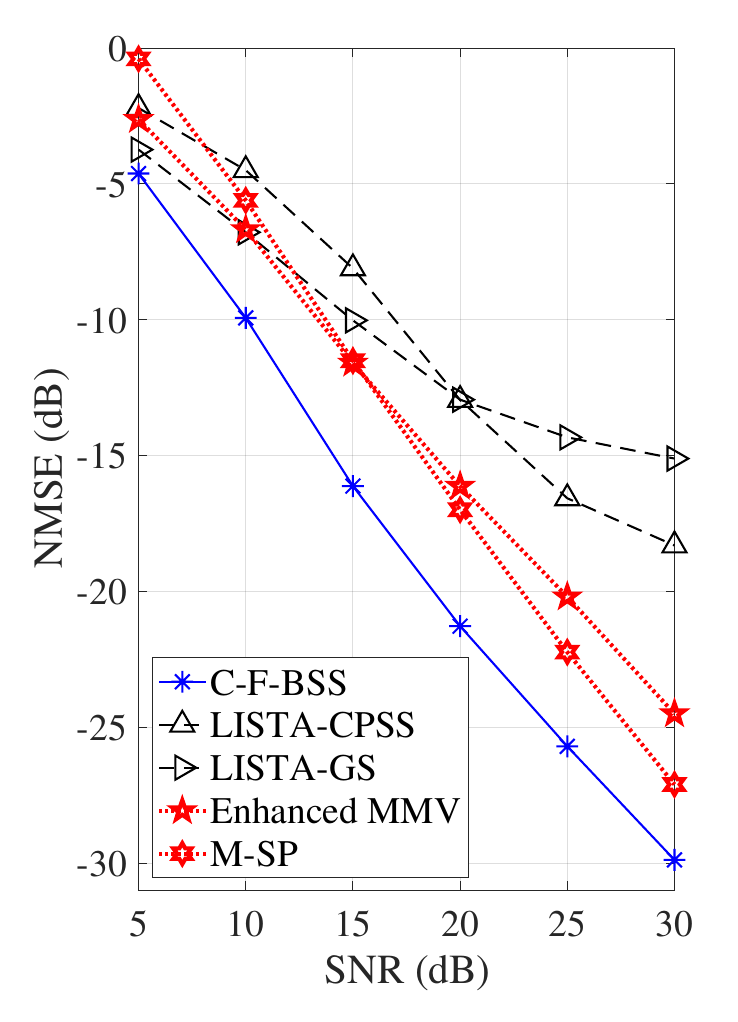}\\
	\end{minipage}
	\hspace{10mm}
	\begin{minipage}[t]{0.19\textwidth}
	  \includegraphics[scale=0.35]{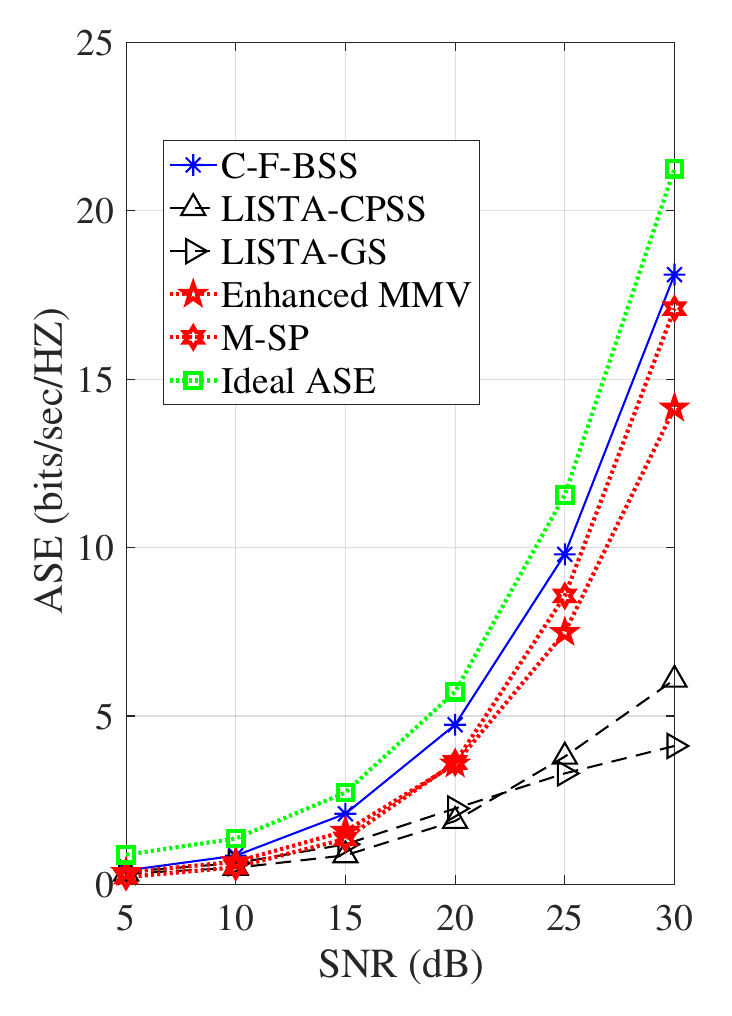}\\
	\end{minipage}
	\vspace{-5mm}
  \caption{NMSE and ASE comparison of estimated channel versus SNR under $\overline{s}=15$, $s_c=10$ and $T=33$.}\label{SNR}
\end{figure}

\begin{table*}
\renewcommand{\arraystretch}{1.1}
\caption{Running time comparison of different schemes under $\overline{s}=15$, $T=33$, $s_c=10$ and $\text{SNR}=30$ dB.}
\label{time consumption}
\centering
\begin{tabular}{|m{25mm}|m{25mm}|m{25mm}|m{25mm}|m{28mm}|m{25mm}|}
\hline
\textbf{Schemes} & \textbf{C-F-BSS} &\textbf{LISTA-GS} &\textbf{LISTA-CPSS} &\textbf{Enhanced MMV} &\textbf{M-SP}\\
\hline
\textbf{Time(s)} & $0.0092$ & $0.0010$ & $0.0014$ & $0.0919$  & $0.0172$ \\
\hline
\end{tabular}
\vspace{-6mm}
\end{table*}

In Fig. \ref{SNR}, NMSE and ASE performance of the proposed scheme versus SNR are demonstrated. Our proposed scheme outperforms other baseline schemes in terms of ASE and NMSE in a wide range of SNR. This phenomenon implies that the accurate estimation of channels of our scheme promotes better spectrum efficiency in massive MIMO system. 

In Table. \ref{time consumption}, the total running time to estimate channels across $L$ frames of all the schemes is demonstrated. Firstly,  our scheme requires less time consumption compared with Enhanced MMV and M-SP, because our scheme requires less computational complexity or fewer number of iterations. Moreover, more time consumption is needed by our scheme compared with LISTA-GS and LISTA-CPSS. These two deep unrolling schemes run faster due to the following reasons: Firstly, they do not require the calculation of selected support set in (\ref{soft thre support selection function}) or exploit the intra-frame sparsity to calculate $l_2$ norm in (\ref{soft thre support selection function}), respectively. Secondly, they estimate the channels among $L$ successive frames in parallel. However, the channel estimation performance of these two deep unrolling schemes is the worst, our scheme achieves better trade-off between channel reconstruction performance and time complexity.
\section*{Conclusion}

In this paper, we develop a two-stage channel estimation scheme which consists of a coarse estimation part and a fine correction part, by jointly exploiting the intra- and inter-frame sparsities.  The proposed channel estimation scheme, namely C-F-BSS, includes two thresholding functions, i.e., BSS and generalized BSS thresholding functions, to greatly improve the channel estimation performance by exploiting intra-frame and small-scale inter-frame sparsities. Meanwhile, with the help of deep unrolling technique, the time complexity of our scheme is reduced compared with traditional optimization-based schemes. The theoretical result is provided to prove the convergence of F-BSS, which also applies for C-BSS by ignoring small-scale inter-frame sparsity. Simulation results are provided to verify that our scheme requires less pilot overhead and achieves better trade-off between channel reconstruction performance and time complexity.

\bibliographystyle{ieeetran}
\bibliography{IEEEabrv, first_confer}

\begin{thebibliography}{10}
\providecommand{\url}[1]{#1}
\csname url@samestyle\endcsname
\providecommand{\newblock}{\relax}
\providecommand{\bibinfo}[2]{#2}
\providecommand{\BIBentrySTDinterwordspacing}{\spaceskip=0pt\relax}
\providecommand{\BIBentryALTinterwordstretchfactor}{4}
\providecommand{\BIBentryALTinterwordspacing}{\spaceskip=\fontdimen2\font plus
\BIBentryALTinterwordstretchfactor\fontdimen3\font minus
  \fontdimen4\font\relax}
\providecommand{\BIBforeignlanguage}[2]{{%
\expandafter\ifx\csname l@#1\endcsname\relax
\typeout{** WARNING: IEEEtran.bst: No hyphenation pattern has been}%
\typeout{** loaded for the language `#1'. Using the pattern for}%
\typeout{** the default language instead.}%
\else
\language=\csname l@#1\endcsname
\fi
#2}}
\providecommand{\BIBdecl}{\relax}
\BIBdecl

\bibitem{58}
Y.~Wang, Z.~Tian, and X.~Cheng, ``Enabling technologies for spectrum and energy
  efficient noma-mmwave-mamimo systems,'' \emph{IEEE Wireless Commun.},
  vol.~27, no.~5, pp. 53--59, Oct. 2020.

\bibitem{5}
A.~Decurninge, L.~G. Ordóñez, and M.~Guillaud, ``Covariance-aided {CSI}
  acquisition with non-orthogonal pilots in massive {MIMO}: A large-system
  performance analysis,'' \emph{IEEE Trans. Inf. Theory}, vol.~66, no.~7, pp.
  4489--4512, Feb. 2020.

\bibitem{41}
Y.~Liao, Y.~Hua, and Y.~Cai, ``Deep learning based channel estimation algorithm
  for fast time-varying {MIMO-OFDM} systems,'' \emph{IEEE Communications
  Letters}, vol.~24, no.~3, pp. 572--576, 2020.

\bibitem{2}
A.~K. Kocharlakota, K.~Upadhya, and S.~A. Vorobyov, ``Impact {of pilot overhead
  and channel estimation on the performance of massive MIMO},'' \emph{IEEE
  Trans. Commun.}, vol.~69, no.~12, pp. 8242--8255, Sep. 2021.

\bibitem{9}
X.~Rao and V.~Lau, ``Compressive sensing with prior support quality information
  and application to massive {MIMO} channel estimation with temporal
  correlation,'' \emph{IEEE Trans. Signal Process.}, vol.~63, no.~18, pp.
  4914--4924, Jun. 2015.

\bibitem{10}
C.-C. Tseng, J.-Y. Wu, and T.-S. Lee, ``Enhanced compressive downlink {CSI}
  recovery for {FDD} massive {MIMO} systems using weighted block ${\ell
  _1}$-minimization,'' \emph{IEEE Trans. Commun.}, vol.~64, no.~3, pp.
  1055--1067, Jan. 2016.

\bibitem{45}
W.~Zhang, M.~Dong, and T.~Kim, ``M{MV}-based sequential {AoA} and {AoD}
  estimation for millimeter wave {MIMO} channels,'' \emph{IEEE Trans. Commun.},
  vol.~70, no.~6, pp. 4063--4077, Apr. 2022.

\bibitem{53}
Y.~Wang, Y.~Zhang, Z.~Tian, G.~Leus, and G.~Zhang, ``Super-resolution channel
  estimation for arbitrary arrays in hybrid millimeter-wave massive {MIMO}
  systems,'' \emph{IEEE J. Sel. Top. Signal Process.}, vol.~13, no.~5, pp.
  947--960, Aug. 2019.

\bibitem{57}
Y.~Zhang, Y.~Wang, Z.~Tian, G.~Leus, and G.~Zhang, ``Efficient angle estimation
  for {MIMO} systems via redundancy reduction representation,'' \emph{IEEE
  Signal Process Lett.}, vol.~29, pp. 1052--1056, Apr. 2022.

\bibitem{56}
W.~Xu, T.~Shen, Y.~Tian, Y.~Wang, and J.~Lin, ``Compressive channel estimation
  exploiting block sparsity in multi-user massive {MIMO} systems,'' in
  \emph{Proc. IEEE Wireless Commun. Networking Conf. (WCNC)}, San Francisco,
  CA, USA, Mar. 2017, pp. 1--5.

\bibitem{50}
F.~Kaltenberger, D.~Gesbert, R.~Knopp, and M.~Kountouris, ``Correlation and
  capacity of measured multi-user {MIMO} channels,'' in \emph{Proc. IEEE Int.
  Symp. Personal, Indoor, Mobile Radio Commun. (PIMRC)}, 2008, pp. 1--5.

\bibitem{55}
L.~Lu, W.~Xu, Y.~Cui, Y.~Dang, and S.~Wang, ``Gamma-distribution-based logit
  weighted block orthogonal matching pursuit for compressed sensing,''
  \emph{Electronics Letters}, vol.~55, no.~17, pp. 959--961, Aug. 2019.

\bibitem{19}
X.~Chen, J.~Liu, Z.~Wang, and W.~Yin, ``Theoretical linear convergence of
  unfolded {ISTA} and its practical weights and thresholds,'' \emph{Proc. Adv.
  Neural Inf. Process. Syst.,}, 2018.

\bibitem{61}
Y.~Shi, H.~Choi, Y.~Shi, and Y.~Zhou, ``Algorithm unrolling for massive access
  via deep neural networks with theoretical guarantee,'' \emph{IEEE
  Transactions on Wireless Communications}, vol.~21, no.~2, pp. 945--959, Feb.
  2022.

\bibitem{20}
Y.~Shi, S.~Xia, Y.~Zhou, and Y.~Shi, ``Sparse signal processing for massive
  device connectivity via deep learning,'' in \emph{Proc. IEEE Int. Conf.
  Commun. Workshops (ICC Workshops)}, Haifa, Israel, Jul. 2020, pp. 1--6.

\bibitem{16}
X.~Ma and Z.~Gao, ``Data-driven deep learning to design pilot and channel
  estimator for massive {MIMO},'' \emph{IEEE Trans. Veh. Technol.}, vol.~69,
  no.~5, pp. 5677--5682, Mar. 2020.

\bibitem{62}
W.~Tong, W.~Xu, F.~Wang, J.~Shang, M.~Pan, and J.~Lin, ``Deep learning
  compressed sensing-based beamspace channel estimation in mmwave massive
  {MIMO} systems,'' \emph{IEEE Wireless Commun. Lett.}, vol.~11, no.~9, pp.
  1935--1939, 2022.

\bibitem{63}
X.~Ma, Z.~Gao, F.~Gao, and M.~Di~Renzo, ``Model-driven deep learning based
  channel estimation and feedback for millimeter-wave massive hybrid {MIMO}
  systems,'' \emph{IEEE J. Sel. Areas Commun.}, vol.~39, no.~8, pp. 2388--2406,
  2021.

\bibitem{21}
Z.~Qin, K.~Scheinberg, and D.~Goldfarb, ``Efficient block-coordinate descent
  algorithms for the group lasso,'' \emph{Math. Program. Comput.}, vol.~5,
  no.~2, pp. 143--169, Mar. 2013.

\bibitem{52}
Y.~Shi, H.~Choi, Y.~Shi, and Y.~Zhou, ``Algorithm unrolling for massive access
  via deep neural networks with theoretical guarantee,'' \emph{IEEE Trans.
  Wireless Commun.}, vol.~21, no.~2, pp. 945--959, Aug. 2022.

\bibitem{64}
E.~Vlachos, G.~C. Alexandropoulos, and J.~Thompson, ``Massive {MIMO} channel
  estimation for millimeter wave systems via matrix completion,'' \emph{IEEE
  Signal Processing Letters}, vol.~25, no.~11, pp. 1675--1679, 2018.

\end{thebibliography}

\end{document}